\documentclass[aps,pra,reprint, amsmath, amssymb,superscriptaddress,nofootinbib]{revtex4-1}

\usepackage{bm}
\usepackage[retainorgcmds]{IEEEtrantools}
\usepackage{graphicx}
\usepackage{mathrsfs}
\usepackage{amsmath}
\usepackage{amssymb}
\usepackage{color}
\usepackage{amsfonts}
\usepackage{times,txfonts}
\usepackage{nicefrac}
\usepackage[colorlinks=true,linkcolor=blue,urlcolor=blue,citecolor=blue,pdfusetitle]{hyperref}
\usepackage{amsmath}

\newcommand{\tr}{\text{tr}}

\begin{document}

\title{Improving the Cram\'{e}r-Rao bound with the detailed fluctuation theorem}
\date{\today}
\author{Domingos S. P. Salazar}
\affiliation{Unidade de Educa\c c\~ao a Dist\^ancia e Tecnologia,
Universidade Federal Rural de Pernambuco,
52171-900 Recife, Pernambuco, Brazil}

\begin{abstract}
In some non-equilibrium systems, the distribution of entropy production $p(\Sigma)$ satisfies the detailed fluctuation theorem (DFT), $p(\Sigma)/p(-\Sigma)=\exp(\Sigma)$. When the distribution $p(\Sigma)$ shows time-dependency, the celebrated Cram\'{e}r-Rao (CR) bound asserts that the mean entropy production rate is upper bounded in terms of the variance of $\Sigma$  and the Fisher information with respect to time. In this letter, we employ the DFT to derive an upper bound for the mean entropy production rate that improves the CR bound. We show that this new bound serves as an accurate approximation for the entropy production rate in the heat exchange problem mediated by a weakly coupled bosonic mode. The bound is saturated for the same setup when mediated by a weakly coupled qubit.

\end{abstract}
\maketitle{}


{\bf \emph{Introduction -}} 
In small-scale thermodynamics, the entropy production $\Sigma$ is usually regarded as a fluctuating quantity \cite{RevModPhys.93.035008,Seifert2012Review,Campisi2011,Esposito2009,Jarzynskia2008,Jarzynski1997,Ciliberto2013,Crooks1998,Gallavotti1995,Evans1993,Hanggi2015,Batalhao2014}. Every time you repeat a thermodynamic process, quantities such as heat, work and entropy production might output different random values. For that reason, it is natural to represent the randomness of the entropy production in a time-dependent probability distribution $p_t(\Sigma)$, where the subscript refers to time, $t \in [0,\infty)$.

Depending on the class of systems, the distribution $p_t(\Sigma)$ might display general properties. Of particular importance is the strong Detailed Fluctuation Theorem (DFT) \cite{Seifert2012Review,Lupos2013}, which is a relation that constrains the asymmetry of $p_t(\Sigma)$,  
\begin{equation}
\label{DFT}
   \frac{p_t(\Sigma)}{p_t(-\Sigma)}=e^{\Sigma},
\end{equation}
forcing positive values of entropy production to be more likely to be observed. The strong DFT (\ref{DFT}) arises, for instance, in driving protocols that are symmetric under time reversal and in the exchange fluctuation framework \cite{Evans2002,Seifert2005,Hasegawa2019,Timpanaro2019B,Merhav2010,Garcia2010,Cleuren2006,Jarzynski2004a,Andrieux2009,Campisi2015}. The most known consequence of (\ref{DFT}) is the integral fluctuation theorem (IFT), $\langle e^{-\Sigma}\rangle_t=1$, which results in the second law of thermodynamics, $\langle \Sigma \rangle_t \geq 0$ for all $t$, from Jensen's inequality, where $\langle f(\Sigma)\rangle_t:=\sum_i f(\Sigma_i)p_t(\Sigma_i)$.

Understanding how the distribution $p_t(\Sigma)$ and the average entropy production $\langle \Sigma \rangle_t$ change over time is important, for instance, for devising optimal thermal machines that operate in finite time. In this context, the Fisher information with respect to time plays a relevant role,
\begin{equation}
\label{Fisher}
I(t):=\langle (\frac{\partial}{\partial t}\log p_t (\Sigma))^2 \rangle_t,
\end{equation}
as it was recently used in stochastic thermodynamics as the intrinsic speed of the
system \cite{Ito2020,Nicholson2020,Hoshino2023}, in the context of thermodynamic length \cite{Crooks2007b} and in connection with the thermodynamic uncertainty relation \cite{Hasegawa2019b,Salazar2022d}. The authors in \cite{Ito2020} noted that the rate of change of the average of any observable is bounded from above by its variance and the temporal Fisher information, evoking the famous Cram\'{e}r-Rao (CR) bound \cite{CramerRao} from estimation theory. Here we are interested in the entropy production as the stochastic quantity, for which the Cram\'{e}r-Rao bound reads
\begin{equation}
\label{CR}
\frac{d\overline{\Sigma}}{dt} \leq \sigma_\Sigma \sqrt{I(t)},
\end{equation}
where $\overline{\Sigma}:=\langle \Sigma \rangle_t$ and $\sigma_\Sigma:=\sqrt{\langle \Sigma^2 \rangle_t - \langle \Sigma \rangle_t^2}$ are both functions of time. Finding upper bounds for the entropy production rate such as (\ref{CR}) is a relevant topic in stochastic \cite{Limkumnerd2017,Dechant2018b,Nishiyama2023} and quantum thermodynamics \cite{PhysRevE.105.L042101}, as they are ultimately related to speed limits \cite{Ito2018,VanVu2021,Yoshimura2021,Shiraishi2018,VanTuan2020,TanVanVu2023,TanVanVu2023b}.

In this letter, we investigate the question: how can the DFT (\ref{DFT}) be used to improve the Cram\'{e}r-Rao upper bound (\ref{CR}) for the entropy production rate? The idea is that, since (\ref{CR}) was derived in the general setting of estimation theory, it might be further improved for the entropy production rate $d \overline{\Sigma}/dt$ in cases where $p_t(\Sigma)$ is constrained by the DFT (\ref{DFT}). Such improvement would have direct impact on the estimation of the entropy production rate in physical systems arbitrarily far from equilibrium.

We show that, in situations where the DFT (\ref{DFT}) is valid, the entropy production rate has an upper bound that improves the CR bound,
\begin{equation}
\label{main}
\frac{d\overline{\Sigma}}{dt} \leq 
\sigma_{h(\Sigma)} \sqrt{I(t)} 
\leq \sigma_\Sigma \sqrt{I(t)},
\end{equation}
which is our main result, where $\sigma_{h(\Sigma)}:=\sqrt{\langle h(\Sigma)^2\rangle - \langle h(\Sigma)\rangle^2}$ and $h(\Sigma):=\Sigma\tanh(\Sigma/2)$. As applications, we also show how the bound acts as a good estimatior for the entropy production rate for the heat exchange problem meadiated by a bosonic mode with Lindblad's dynamics in comparison with the CR bound (\ref{CR}). We also show how the bound is saturated for the same problem when mediated by a qubit. We argue that the behavior of the bound in those cases is not accidental: the bound (\ref{main}) is always saturated for a time dependent maximal distribution \cite{Salazar2021}, which was originally derived as the distribution that maximizes Shannon's entropy for a given mean, while satisfying the DFT (\ref{DFT}). We show that the qubit case falls in the maximal distribution family and the bosonic case is very close to it.

{\bf \emph{Formalism -}}
Let $\Sigma \in S=\{\Sigma_1, \Sigma_2,...\}$ be a random variable with distribution $p_t(\Sigma)$ that depends on time $t\in [0,\infty)$ and satisfies the DFT (\ref{DFT}). Let $\phi(\Sigma)$ be any odd function,
\begin{equation}
\label{form1}
\phi(-\Sigma)=-\phi(\Sigma).
\end{equation}
The DFT imposes the following known property \cite{Hasegawa2019,Merhav2010} on the average of odd functions,
\begin{equation}
\label{form2}
\langle \phi(\Sigma) \rangle_t = \langle \phi(\Sigma)\tanh(\frac{\Sigma}{2})\rangle_t.
\end{equation}
The time derivative of (\ref{form2}) yields
\begin{equation}
\label{form3}
\frac{d}{dt}\langle \phi(\Sigma) \rangle_t = \sum_i \phi(\Sigma_i)\tanh(\frac{\Sigma_i}{2}) \frac{\partial}{\partial t}p_t(\Sigma_i),
\end{equation}
which can be written as
\begin{equation}
\label{form4}
\frac{d}{dt}\langle \phi(\Sigma) \rangle_t = \sum_i \sqrt{p_t(\Sigma_i)}[\phi(\Sigma_i)\tanh(\frac{\Sigma_i}{2})-c] \frac{\dot{p}_t(\Sigma_i)}{\sqrt{p_t(\Sigma_i)}},
\end{equation}
where $c$ is any constant and $\dot{p_t}(\Sigma_i):=\partial p_t (\Sigma_i)/\partial t$. Now using Cauchy–Schwarz inequality, one obtains from (\ref{form4})
\begin{equation}
\label{form5}
\big(\frac{d}{dt}\langle \phi(\Sigma)\rangle_t\big)^2 \leq \langle [\phi(\Sigma_i)\tanh(\frac{\Sigma_i}{2})-c]^2 \rangle_t I(t),
\end{equation}
with the Fisher information $I(t)$ given by (\ref{Fisher}). Finally, considering the special case $\phi(\Sigma)=\Sigma$ and setting $c=\langle \Sigma \tanh(\Sigma/2) \rangle_t =\langle h(\Sigma)\rangle_t$, we obtain the first inequality in (\ref{main}),
\begin{equation}
\label{form6}
|\frac{d\overline{\Sigma}}{dt}| \leq 
\sigma_{h(\Sigma)} \sqrt{I(t)},
\end{equation}
note that one could write $|d\overline{\Sigma}/dt| = d\overline{\Sigma}/dt$ by construction, since the DFT (and the second law, $\langle \Sigma \rangle_t \geq 0$) works for all time $t>0$. The second inequality in (\ref{main}) follows from
\begin{equation}
\label{form7}
\tanh(\Sigma/2)^2 \leq 1 \rightarrow h(\Sigma)^2 \leq \Sigma^2,
\end{equation}
which, upon taking the average of (\ref{form7}) over $p_t(\Sigma)$ and subtracting $\langle h(\Sigma)\rangle^2$, it yields
\begin{equation}
\label{form8}
\langle h(\Sigma)^2 \rangle_t -\langle h(\Sigma) \rangle_t^ 2  \leq \langle \Sigma^2 \rangle_t -\langle \Sigma \rangle_t^2,
\end{equation}
where we used $\langle h(\Sigma)\rangle_t=\langle \Sigma \rangle_t$ from (\ref{form2}). Finally, we have from (\ref{form8}),
\begin{equation}
\sigma_{h(\Sigma)} \sqrt{I(t)} \leq \sigma_\Sigma \sqrt{I(t)},
\end{equation}
which is the second inequality of our main result (\ref{main}), showing that it improves the CR bound. In  the examples below, we start with a dynamics that allows one to compute both $d\overline{\Sigma}/dt$ and $p_t(\Sigma)$ exactly. We check that $p_t(\Sigma)$ satisfy the DFT (\ref{DFT}), then we use $p_t(\Sigma)$ to find $\sigma_\Sigma$, $\sigma_{h(\Sigma)}$ and $I(t)$. Finally, we show the bounds (\ref{main}) as a function of time in Figs 1 and 2. Then, we discuss why the bound is a surprisingly good approximation for $d\overline{\Sigma}/dt$ in both cases.

{\bf \emph{Application I: bosonic mode-}} We consider a bosonic mode with Hamiltonian $H=\hbar\omega (a^\dagger a+1/2)$ weakly coupled to a thermal reservoir so that the system satisfies a Lindblad's equation \cite{Santos2017a,Salazar2019,Denzler2019},
\begin{equation}
\label{Lind}
\partial_t \rho = \frac{-i}{\hbar}[H,\rho] + D(\rho),
\end{equation}
for the dissipator given by
\begin{equation}
D(\rho)=\gamma(\overline{n}_2+1)[a\rho a^\dagger - \frac{1}{2}\{a^\dagger a, \rho\}]+\gamma \overline{n}_2[a^\dagger \rho a -\frac{1}{2}\{a a^\dagger , \rho\}],
\end{equation}

\begin{figure}[htp]
\includegraphics[width=3.3 in]{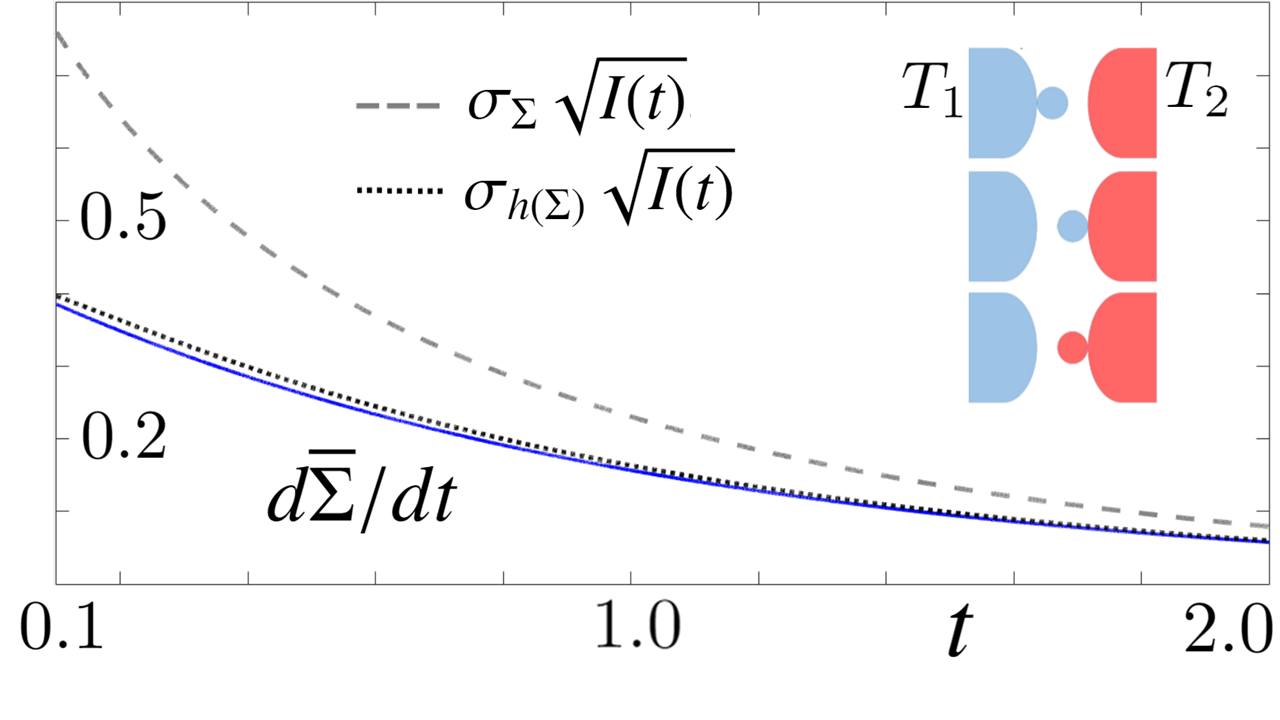}
\caption{(Color online) Entropy production rate $d\overline{\Sigma}/dt$ (blue line) as a function of time for the heat exchange problem mediated by a bosonic mode ($\gamma=1$, $\hbar \omega/k_B T_1 = 1$, $T_2=T_1/2$). The Cram\'{e}r-Rao bound $\sigma_\Sigma \sqrt{I(t)}$ is depicted in dashed gray and $\sigma_{h{\Sigma}} \sqrt{I(t)}$ is the dotted black line. In this case, the entropy production rate is very close to the bound, but does not saturate it.}
\label{fig1}
\end{figure}
where $\gamma$ is a constant, $\overline{n}_i=[\exp(\hbar \omega/k_BT_i)-1]^{-1}$ and $\beta_i=1/(k_B T_i)$, $i\in \{1,2\}$. 
The system is prepared in thermal equilibrium with the first reservoir (temperature $T_1$). At $t=0$, an energy measurement is performed, resulting in $E_0 = \hbar \omega (n_0+1/2)$, $n_0 \in\{0,1,2...\}$. Then, for $t>0$, the system is placed in thermal equilibrium with a second reservoir (temperature $T_2)$ with dynamics (\ref{Lind}). At a given $t>0$, a second measurement is performed, resulting in $E_t = \hbar \omega (n_t+1/2)$, where $n_t \in \{0,1,2...\}$, where $n_t := \tr(a^\dagger a \rho_t)$

The time dependent random variable $\Sigma := -(\beta_2-\beta_1)(E_t - E_0)$ \cite{Campisi2015,Timpanaro2019B,Sinitsyn2011} is the entropy production in this case: intuitively, you could see the bosonic mode as part of the first reservoir, so that the second reservoir transfers $\Delta E$ to it in the form of heat. This heat exchange results in the entropy flux $-\beta_2 \Delta E$ in the second reservoir and $\beta_1 \Delta E$ in the first reservoir, which results in a total entropy flux $\Phi=-(\beta_2-\beta_1)(E_t - E_0)$. The total entropy variation (system + reservoirs) is zero, $\Delta S = \Sigma - \Phi=0$, so the flux $\Phi$ must be compensated with the entropy production $\Sigma=\Phi=-(\beta_2-\beta_1)(E_t - E_0)$. 

For simplicity, let us consider $\hbar \omega / k_B T_1 = 1$ (hot) and $T_2=T_1/2 < T_1$ (cold), such that $\Delta \beta \hbar \omega = 1$. The average entropy production (over all possible $n_0$ and $n_t$) is given by
\begin{equation}
\label{EPbosonic}
\langle \Sigma \rangle_t = (\overline{n}_1-\overline{n}_2)(1-e^{-\gamma t}),
\end{equation}
from the dynamics (\ref{Lind}) directly. The distribution $p_t(\Sigma)$ has a closed form \cite{Salazar2019,Denzler2019} that satisfy the DFT (\ref{DFT}),
\begin{equation}
\label{PsigmaHO}
    p_t(\Sigma)=\frac{1}{A(\lambda_t)} \exp(\frac{\Sigma}{2}-\lambda_t \frac{|\Sigma|}{2}),
\end{equation}
with support $\Sigma \in \{\pm m\}$, $m=0,1,2,..$ and the normalization constant reads $A(\lambda_t):=[1-\exp(-1/2-\lambda_t/2)]^{-1}+[1-\exp(1/2-\lambda_t/2)]^{-1}-1$, where $\lambda_t > 1$. This situations corresponds to the heat exchange of $\Delta E$ from a cold ($T_2$) to a hot ($T_1$) reservoir mediated by a bosonic mode, so that the second law $\langle \Sigma \rangle_t \geq 0 \rightarrow 
\langle \Delta E \rangle_t \leq 0$ is telling that the energy should flow from the hot to the cold reservoir on average, as expected. The value of $\lambda_t$ is given implicitly by
\begin{equation}
\label{EPratebosonic}
 (\overline{n}_1-\overline{n}_2)(1-e^{-\gamma t}) = \frac{e^{-1/2+\lambda_t/2}}{(e^{-1/2+\lambda_t/2}-1)^2}-\frac{e^{-1/2-\lambda_t/2}}{(e^{-1/2-\lambda_t/2}-1)^2},
\end{equation}
where rhs comes from the distribution (\ref{PsigmaHO}) and the lhs comes from (\ref{EPbosonic}). Using $\lambda_t$ in (\ref{PsigmaHO}), one calculates the Fisher information using (\ref{Fisher}). The values $\sigma_\Sigma$ and $\sigma_{h(\Sigma)}$ are also given in terms of (\ref{PsigmaHO}), using $\langle \Sigma^2 \rangle_t = \sum_i \Sigma_i^2 p_t(\Sigma_i)$ and $\langle h(\Sigma_i)^2 \rangle_t = \sum_i h(\Sigma_i)^2 p_t(\Sigma_i)$.

In Fig.1, we plot the entropy production rate $d\overline{\Sigma}/dt$ from (\ref{EPbosonic}) as a function of time for $\hbar \omega / k_B T_1 = 1$, $T_2=T_1/2 < T_1$ and $\gamma=1$. We also plot the Cram\'{e}r-Rao
upper bound (\ref{CR}) and our result (\ref{main}) for comparison, showing that the proposed bound is actually a good approximation to the entropy production rate when compared to the CR bound. In the discussion subsection, we will provide some intuition about the reason for such good approximation.

{\bf \emph{Application  II: qubit -}} We consider the same measurement scheme as before, the only difference is that the system mediating the heat exchange is a qubit with Hamiltonian $H=\hbar \omega \hat{\sigma}^\dagger \hat{\sigma}$, where $\hat{\sigma}^\dagger=|1\rangle \langle 0|$ and $\hat{\sigma}=|0\rangle \langle 1|$. The systems evolves with a Lindblad's dynamics (\ref{Lind}) with dissipator
\begin{equation}
\label{dissipatorqubit}
    D_2(\rho)=\gamma(1-\overline{n}_2)(\hat{\sigma} \rho \hat{\sigma}^\dagger-\frac{1}{2}\{\hat{\sigma}^\dagger\hat{\sigma},\rho\})+\gamma \overline{n}_2(\hat{\sigma}^\dagger\rho \hat{\sigma}-\frac{1}{2}\{\hat{\sigma} \hat{\sigma}^\dagger,\rho\}),
\end{equation}
where $\overline{n}_i=1/(1+e^{-\beta \omega})$ is the thermal occupation for this case. As in the previous example, the qubit is prepared in thermal equilibrium with the first reservoir ($T_1$) and, at $t=0$ the first energy measurement takes place ($E_0 = \hbar \omega n_0$, $n_0 \in \{0,1\}$), after that it is placed in thermal contact with the second reservoir ($T_2$) for a time $t>0$, modelled with the dynamics (\ref{Lind}), when a second measurement takes place $(E_t = \hbar \omega n_t$, $n_t \in \{0,1\})$. Using the same reasoning as before, the entropy production is given by $\Sigma=-\hbar \omega\Delta \beta (n_t - n_0)$, where $n_0$ and $n_t$ are Bernoulli random variables. Again, for simplicity, let us consider $\hbar \omega / k_B T_1 = 1$ (hot) and $T_2=T_1/2 < T_1$ (cold), such that $\Delta \beta \hbar \omega = 1$. The average entropy production over all possible $n_0$ and $n_1$ yields from the dynamics
\begin{equation}
\langle \Sigma \rangle_t = (\overline{n}_1 - \overline{n}_2)(1-e^{-\gamma t}),
\end{equation}
just as before (\ref{EPbosonic}), but now the occupation numbers $\overline{n}_i$ have different values, $\overline{n}_1=1/(1+e)$, $\overline{n}_2=1/(1+e^2)$.
For this setup, one has $P(n_0=1)=\overline{n}_1$ and $P(n_t=1|n_0) = \overline{n}_2 + (n_0 - \overline{n_2})\exp(-\gamma t)$. Considering all possibilities of $n_0$ and $n_t$ results in the following distribution for $p_t(\Sigma)$
\begin{equation}
\label{quibitpdf0}
p_t(0)=1-(1-e^{-\gamma t})(\overline{n}_1+\overline{n}_2-2\overline{n}_1 \overline{n}_2),
\end{equation}
for $\Sigma=0$ and
\begin{equation}
\label{qubitpdf}
p_t(\Sigma)=(1-p_t(0))\frac{e^{\Sigma/2}}{e^{1/2}+e^{-1/2}},
\end{equation}
for $\Sigma \in \{\pm 1 \}$, which satisfy the DFT (\ref{DFT}).

\begin{figure}[htp]
\includegraphics[width=3.3 in]{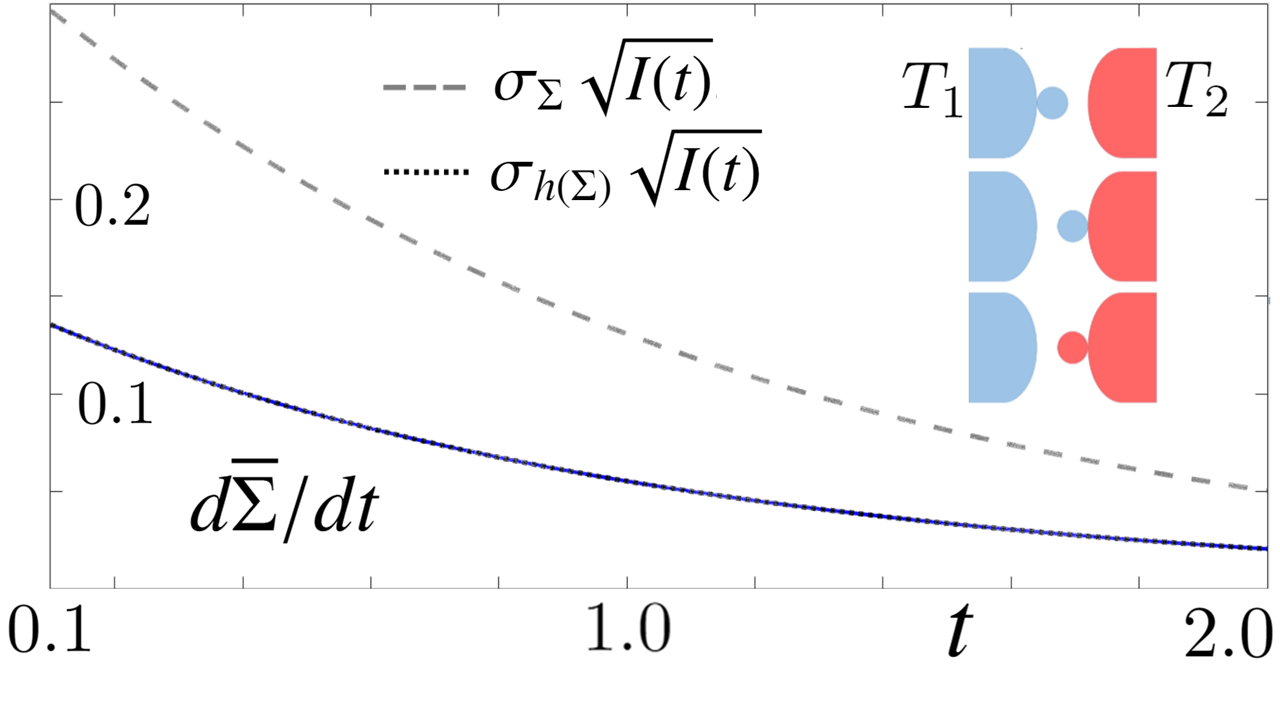}
\caption{(Color online) Entropy production rate $d\overline{\Sigma}/dt$ (blue line) as a function of time for the heat exchange problem mediated by a qubit ($\gamma=1$, $\hbar \omega/k_B T_1 = 1$, $T_2=T_1/2$). The Cram\'{e}r-Rao upper bound $\sigma_\Sigma \sqrt{I(t)}$ is depicted in dashed gray and $\sigma_{h{\Sigma}} \sqrt{I(t)}$ is the dotted black line. In this case, the entropy production rate matches the upper bound. Actually, $p_t(\Sigma)$ is a particular case of the maximal distribution, which always saturates the bound.}
\label{fig1}
\end{figure}
Using the distribution (\ref{qubitpdf}), we compute $\langle \Sigma \rangle_t = \tanh(1/2)(1-p_t(0))$, $\langle \Sigma ^2 \rangle_t = 1-p_t(0)$, $\langle h(\Sigma )^2\rangle_t = \tanh^2(1/2)(1-p_t(0))$, which allows one to write $\sigma_\Sigma$ and $\sigma_{h(\Sigma)}$ as functions of time. Finally, one can use (\ref{qubitpdf}) and (\ref{quibitpdf0}) to find the Fisher information (\ref{Fisher}) also as a function of time. In this particular case, it yields $I(t)=(\partial p_t(0)/\partial t)^2/[p_t(0)(1-p_t(0))]$.

In Fig.2, as in the previous example, we plot the entropy production rate, also given by $d\overline{\Sigma}/dt = (\overline{n}_1-\overline{n}_2)e^{-\gamma t}$, as a function of time for $\hbar \omega / k_B T_1 = 1$, $T_2=T_1/2 < T_1$ and $\gamma=1$. We also plot the Cram\'{e}r-Rao
bound (\ref{CR}) and our result (\ref{main}). In the case of the qubit, the upper bound is saturated while the CR bound is not. This fact led us to investigate what would be a sufficient condition for the saturation, as discussed in the next section.

{\bf \emph{Discussion -}} We note the bound (\ref{main}) was verified in both applications, as expected from the DFT (\ref{DFT}). However, the bound also worked as a good estimator of the entropy production rate in both cases, matching the exact value for the qubit. Now we investigate the intuition behind it. First, we consider the following a general distribution $p_t(\Sigma)$ of the exponential family \cite{Crooks2007b,Ito2020} that satisfies the DFT (\ref{DFT}),
\begin{equation}
\label{expofam}
p_t(\Sigma)= \frac{1}{Z(\lambda_t)}\exp(\frac{\Sigma}{2} - \frac{\lambda_t}{2} f(\Sigma))),
\end{equation}
where $f$ is even, $f(\Sigma)=f(-\Sigma)$ and $Z(\lambda_t):=\int \exp(\Sigma/2 - (\lambda_t/2)f(\Sigma))$ is a normalization constant
. We also have $- \partial \log Z(\lambda_t) / \partial \lambda_t = (1/2)\langle f(\Sigma)\rangle_t$ and
\begin{equation}
\label{EPrateexpo}
    \frac{d\overline{\Sigma}}{dt}=\frac{\dot{\lambda_t}}{2}(\langle \Sigma \rangle_t \langle f(\Sigma)\rangle_t - \langle \Sigma f(\Sigma) \rangle_t)).
\end{equation}
Now using property (\ref{form2}) in (\ref{EPrateexpo}) for the odd functions $\phi(\Sigma)=\Sigma$ and $\phi(\Sigma)=\Sigma f(\Sigma)$, we have $\langle \Sigma \rangle_t = \langle h(\Sigma) \rangle_t$ and $\langle \Sigma f(\Sigma)\rangle_t = \langle h(\Sigma)f(\Sigma)\rangle_t$, which results in 
\begin{equation}
\label{EPrateexpo2}
    \frac{d\overline{\Sigma}}{dt}=\frac{\dot{\lambda_t}}{2}(\langle h(\Sigma) \rangle_t \langle f(\Sigma)\rangle_t - \langle h(\Sigma) f(\Sigma) \rangle_t)).
\end{equation}
The Fisher information (\ref{Fisher}) from (\ref{expofam}) is given by

\begin{equation}
\label{Fisherexpo}
I(t)=\langle [f(\Sigma) - \langle f(\Sigma)\rangle_t]^2 \rangle_t \frac{\dot{\lambda_t}^2}{4
} = \sigma_{f(\Sigma)}^2 \frac{\dot{\lambda_t}^2}{4
}.
\end{equation}
Finally, using (\ref{EPrateexpo2}) and (\ref{Fisherexpo}) in (\ref{form6}), we obtain for $\dot{\lambda}_t \neq 0$
\begin{equation}
| \langle h(\Sigma) f(\Sigma) \rangle_t - \langle h(\Sigma) \rangle_t \langle f(\Sigma)\rangle_t |\leq \sigma_{h(\Sigma)}\sigma_{f(\Sigma)}, 
\end{equation}
which can be rearranged as
\begin{equation}
\label{Pearsons}
|r_{h(\Sigma),f(\Sigma)}|:=\frac{|cov_{h(\Sigma),f(\Sigma)}|}{\sigma_{h(\Sigma)} \sigma_{f(\Sigma)}} \leq 1.
\end{equation}
The lhs of (\ref{Pearsons}) is the absolute value of the Pearson correlation coefficient $r_{h(\Sigma),f(\Sigma)}$ between the random variables $h(\Sigma)$ and $f(\Sigma)$. The saturation of the bound is thus obtained for $|r_{h(\Sigma),f(\Sigma)}|=1$, resulting from the identity $f(\Sigma) = h(\Sigma)$. In this case, the pdf (\ref{expofam}) reads
\begin{equation}
\label{maximal}
p_t(\Sigma)= \frac{1}{Z(\lambda_t)}\exp(\frac{\Sigma}{2} - \lambda_t \frac{\Sigma}{2} \tanh(\frac{\Sigma}{2})),
\end{equation}
which is the maximal distribution \cite{Salazar2021}, originally derived as the distribution that maximizes Shannon's entropy for a given mean with the DFT (\ref{DFT}) as a constraint.

Comparing the maximal distribution (\ref{maximal}) with the qubit example (\ref{qubitpdf}) shows that it is indeed a member of this family (for a specific support $\Sigma \in \{-\hbar \omega \Delta \beta ,0,\hbar \omega \Delta \beta\})$. For that reason, the entropy production rate actually matches the upper bound in Fig.~2. Alternatively, the bosonic case (\ref{PsigmaHO}) does not saturate the bound in Fig.~1, but it is very close. Using the notation (\ref{expofam}), the bosonic case has $f(\Sigma)=|\Sigma| \approx h(\Sigma)$, which is a close approximation for the maximal distribution. In general, for any given system in the exponential family, the upper bound will serve as a good approximation whenever $|r_{f(\Sigma),h(\Sigma)}| \approx 1$.

{\bf \emph{Conclusions -}}
We used the DFT (\ref{DFT}) improve the Cram\'{e}r-Rao upper bound for the entropy production rate. We checked the behavior of the bound in the heat exchange problem mediated by two relevant physical systems in the weak couling approximation: a bosonic mode and a qubit. We found that the bound is very close to the entropy production rate as a function of time for the bosonic case and it saturates for the qubit. Finally, in a more general setting, we showed that the bound is actually saturated for a maximal distribution, which contains the qubit example as a particular case and it approximates the bosonic case. Due to the recent developments of the DFT outside stochastic thermodynamics, specially in quantum correlated systems [Ref], we believe this result will have impact in the understanding of the limiting behavior of open quantum systems.

\bibliography{lib7}
\end{document}